\documentclass[aps,prb,preprint,superscriptaddress]{revtex4-1}
\usepackage{amssymb}
\usepackage{amsmath}
\usepackage{color}
\usepackage{bm}
\usepackage{epsfig}
\usepackage{graphicx}
\usepackage{dcolumn}
\usepackage[pagewise,columnwise]{lineno}
\usepackage{setspace}
\usepackage[utf8]{inputenc}
\usepackage{color}
\usepackage{nicefrac}
\usepackage{hyperref}

\begin{document}


\title{Magnetic properties of Gd$T_2$Zn$_{20}$ (T = Fe, Co) investigated by X-ray diffraction and spectroscopy}

\author{J. R. L. Mardegan}
\affiliation{Instituto de F\'isica ``Gleb Wataghin," Universidade Estadual de Campinas, Campinas, S\~ao Paulo 13083-859  Brazil}
\affiliation{Deutsches Elektronen-Synchrotron (DESY), Hamburg 22607, Germany}

\author{G. Fabbris}
\affiliation{Advanced Photon Source, Argonne National Laboratory, Argonne, Illinois 60439 USA}
\affiliation{Department of Physics, Washington University, St. Louis, Missouri 63130 USA}
\affiliation{Department of Condensed Matter Physics and Materials Science, Brookhaven National Laboratory, Upton, New York 11973, USA}

\author{L. S. I. Veiga}
\affiliation{Deutsches Elektronen-Synchrotron (DESY), Hamburg 22607, Germany}

\author{S. Francoual}
\affiliation{Deutsches Elektronen-Synchrotron (DESY), Hamburg 22607, Germany}

\author{J. Strempfer}
\affiliation{Deutsches Elektronen-Synchrotron (DESY), Hamburg 22607, Germany}

\author{D. Haskel}
\affiliation{Advanced Photon Source, Argonne National Laboratory, Argonne, Illinois 60439 USA}

\author{R. A. Ribeiro}
\affiliation{CCNH, Universidade Federal do ABC (UFABC), Santo Andr\'e, S\~ao Paulo 09210-580 Brazil}

\author{M. A. Avila}
\affiliation{CCNH, Universidade Federal do ABC (UFABC), Santo Andr\'e, S\~ao Paulo 09210-580 Brazil}

\author{C. Giles}
\affiliation{Instituto de F\'isica ``Gleb Wataghin," Universidade Estadual de Campinas, Campinas, S\~ao Paulo 13083-859 Brazil}

\date{\today}


\begin{abstract} 
We investigate the magnetic and electronic properties of the Gd$T_2$Zn$_{20}$ ($T$ = Fe and Co) compounds using X-ray resonant magnetic scattering (XRMS), X-ray absorption near-edge structure (XANES) and X-ray magnetic circular dichroism (XMCD) techniques. 
The XRMS measurements reveal that the GdCo$_2$Zn$_{20}$ compound has a commensurate antiferromagnetic spin structure with a magnetic propagation vector $\vec{\tau}$ = $(\frac{1}{2},\frac{1}{2},\frac{1}{2})$ below the N\'eel temperature ($T_N \sim$ 5.7~K). Only the Gd ions carry a magnetic moment forming an antiferromagnetic structure with magnetic representation $\Gamma_6$. 
For the ferromagnetic GdFe$_2$Zn$_{20}$ compound, an extensive investigation was performed at low temperature and under magnetic field using XANES and XMCD techniques. 
A strong XMCD signal of about 12.5 $\%$ and 9.7 $\%$ is observed below the Curie temperature ($T_C \sim$ 85 K) at the Gd-$L_2$ and $L_3$ edges, respectively. 
In addition, a small magnetic signal of about 0.06~$\%$ of the jump is recorded at the Zn $K$-edge suggesting that the Zn 4$p$ states are spin polarized by the Gd 5$d$ extended orbitals.

\end{abstract}



\maketitle


\section{INTRODUCTION}

\begin{figure}[!b]
\centering
\includegraphics[trim=0.0cm 0.00cm 0.0cm 0.0cm, clip=true, totalheight=0.2 \textheight, angle=0]{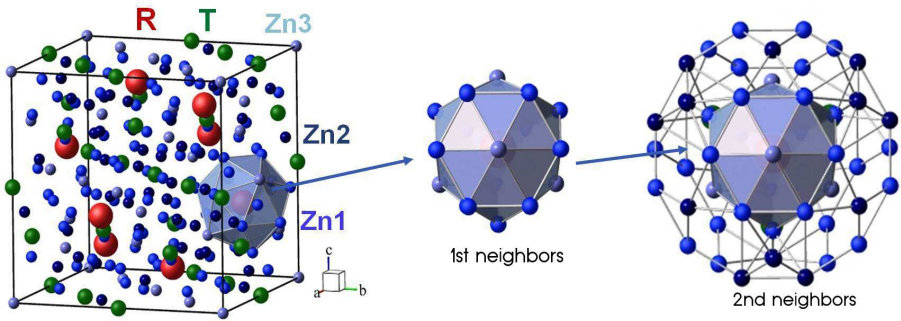}
\caption{(Color online) $RT_2$Zn$_{20}$ structural representation. The Zn ions are shown at the three sites (label Zn1-3) and the Zn cages are shown in detail in which there is a $R$ atom inside. }
\label{fig:fig0}
\end{figure}

The demand for new materials with interesting and useful physical properties has led to a fast development in material science.
Properties such as superconductivity, magnetic ordering, nearly ferromagnetic Fermi-liquid and heavy fermion behavior have been observed in many materials, including the family of complex intermetallic compounds $RT_2$Zn$_{20}$ ($R$ = rare earth, $T$ = transition metal).\cite{Torikachvili_PNAS_2007,Jia_Nature_2007,Jia_PRB_2008, Jia_PRB_2009, onimaru2010superconductivity} 
This family, first reported two decades ago by Nasch $et~al$.\cite{nasch1997ternary}, has been extensively used as a model system due to its rather unique structure which features a complex but well ordered crystal structure. 
The $RT_2$Zn$_{20}$ compounds have a cubic structure with $F d \bar{3} m$ (No. 227) space group in which the $R$ and $T$ ions occupy the crystallographic sites 8$a$ and 16$d$, respectively. 
Moreover, these ions are each encapsulated in quasi-spherical cages formed exclusively by Zn ions, which occupy three inequivalent Wyckoff positions given by 16$c$, 48$f$, and 96$g$ as can be viewed in Fig.~\ref{fig:fig0}.  
Two particular properties observed in these materials have attracted great attention: a remarkably high magnetic ordering temperature observed in the $R$Fe$_2$Zn$_{20}$ series (although it contains less than $5\%$ of $R$ ion) and a nearly ferromagnetic Fermi-liquid behavior in YFe$_2$Zn$_{20}$.\cite{Jia_Nature_2007,Jia_PRB_2009}

Several works have argued, based on macroscopic measurements and band structure calculations, that the elevated magnetic ordering temperatures and the type of magnetic ordering are attributed to a high density of states (DOS) at the Fermi level.\cite{Jia_Nature_2007,Jia_PRB_2008, Jia_PRB_2009, tian2010magnetic, Ni_PRB_2011} 
For instance, the Co based compounds GdCo$_2$Zn$_{20}$ and TbCo$_2$Zn$_{20}$ present an antiferromagnetic (AFM) ordering below $\sim$ 5.7~K and 2.5~K, respectively, in which the larger $R$-$R$ distance ($\sim$ 6 \AA{}) supports an indirect interaction and a low transition temperature. 
When the Co ions are replaced by Fe ions (GdFe$_2$Zn$_{20}$ and TbFe$_2$Zn$_{20}$) the compounds exhibit ferromagnetic (FM) ordering with the transition temperatures drastically raised to 86~K and 66~K, respectively. 
The relatively long distance between rare earth ions in the structure weakens the Ruderman-Kittel-Kasuya-Yosida (RKKY) exchange interaction, while the high density of Fe 3$d$ bands at the Fermi level directly affects the conduction electrons. 

Recently, a detailed study of the magnetic structure of TbCo$_2$Zn$_{20}$ and TbFe$_2$Zn$_{20}$ through magnetic neutron scattering at low temperature was reported.\cite{tian2010magnetic} 
Although the Tb-based compounds present similar magnetic properties as compared to GdCo$_2$Zn$_{20}$ and GdFe$_2$Zn$_{20}$, the absence of crystalline electric field (CEF) at first order and a strong RKKY interaction in the Gd-compounds affect the electronic and magnetic interactions between the rare earth ions and the surrounding matrix. 
In particular, the investigation of the compounds with half-filled 4$f^7$ shell (Gd-based materials) at low temperature can provide information about the physical properties and are very important as reference compounds due to their lack of orbital momentum, i.e., $S = 7/2$ and $L = 0$, which leads to magnetic properties that are unaffected by spin-orbit coupling. 
In order to understand the implications of such interactions we have used spectroscopy and magnetic scattering techniques to probe in detail the electronic and magnetic properties of the Gd$T_2$Zn$_{20}$ family. 
The large Gd neutron absorption cross section leaves X-ray measurements as the ideal option to probe of the magnetic and electronic properties in these materials. 
Furthermore, the incoming beam energy can be tuned to the absorption edge, thus providing chemical and atomic selectivity, i.e., the magnetic response of each element can be probed separately.

Here we report the magnetic and electronic structure at low temperature of the Gd$T_2$Zn$_{20}$ ($T$ = Fe, Co) compounds using X-ray resonant magnetic scattering (XRMS), X-ray absorption near-edge structure (XANES) and X-ray magnetic circular dichroism (XMCD) techniques. 
The XRMS measurements performed on GdCo$_2$Zn$_{20}$ reveal a commensurate antiferromagnetic ordering with a magnetic propagation vector $\vec{\tau} = (\frac{1}{2},\frac{1}{2},\frac{1}{2})$ below $T_N = 5.72(6)$ K in which the Gd magnetic moments are aligned following the magnetic representation $\Gamma_6$. 
The XMCD measurements performed below the Curie Temperature ($T_C = 85(2)$ K) in GdFe$_2$Zn$_{20}$ display a dichroic signal of 12.5~$\%$ and 9.7~$\%$ of the absorption jump for Gd $L_{2}$ and $L_3$ edges, respectively. 
Surprisingly, a magnetic signal of about 0.06~$\%$ is detected at the Zn $K$-edge which suggests that the Zn ions are spin polarized. 
This magnetic signal might originate from the hybridization between the extended Gd 5$d$ bands with the empty Zn 4$sp$ states. 
Absorption measurements performed at the Fe $K$-edge do not reveal any magnetic contribution coming from the iron ions above the background level. 


\section{EXPERIMENTAL DETAILS} 

High quality single crystals of GdFe$_2$Zn$_{20}$ and GdCo$_2$Zn$_{20}$ were grown at UFABC by Zn self flux method\cite{Canfield_PhilMag_1992,Ribeiro_PhilMag_2012} similar to that reported in previous studies on the family.\cite{Jia_Nature_2007, Jia_PRB_2008,  Jia_PRB_2009}
In order to perform the absorption measurements at the Gd $L_{2,3}$, Fe and Zn $K$ edge, selected single crystals of GdFe$_2$Zn$_{20}$ were ground and sieved, resulting in fine powders with grain sizes around $3$-$5~\mu m$. 
The magnetic diffraction measurements were done on a high quality GdCo$_2$Zn$_{20}$ single crystal cut to dimensions of approximately 2 x 2 x 0.5 mm$^3$. 
The crystalline piece was carefully polished to achieve a flat surface perpendicular to the $[111]$ direction, yielding a mosaic width of approximately 0.02$^{\circ}$. 
The phase purity of the samples was confirmed by powder diffraction using conventional laboratory X-ray sources. 
Temperature dependent magnetic susceptibility measurements (not shown here) were performed using a commercial superconducting quantum interference device (Quantum Design MPMS-SQUID) to verify the magnetic ordering temperatures, the effective magnetic moments and the Curie-Weiss constants.

The resonant diffraction measurements were performed at beamline 6-ID-B at the Advanced Photon Source (APS), Argonne National Laboratory (Argonne, IL/USA), whereas the absorption measurements were conducted at 4-ID-D (APS)\cite{Freeland_RSI_2002} and at beamline P09 at PETRA III (DESY, Hamburg/Germany).\cite{Strempfer_JSR_2013}

\subsection{Absorption experiment}
XANES and XMCD spectra obtained at low temperature for the Gd $L_{2,3}$, Fe and Zn $K$-absorption edges were performed in transmission geometry on powdered GdFe$_2$Zn$_{20}$ samples. 
The samples were cooled down by a displex cryostat with base temperature around 7~K. 
XMCD spectra were performed in helicity switching mode in which the left and right circular polarization was obtained by means of diamond phase plates.\cite{Hirano_JJAP_1991} 
The degree of circularly polarized beam was higher than 95~$\%$ for both beamlines (P09 and 4-ID-D).\cite{Francoual_JPCS_2013, Lang_RSI_1995} 
An external magnetic field of H = 2.0~T (at APS) and 0.8~T (at DESY) was applied in the GdFe$_2$Zn$_{20}$ samples along and opposite to the incident beam wave vector $\hat{k}$ to align the ferromagnetic domains and to correct for non-magnetic artifacts in the XMCD data. 
Those external magnetic fields were enough to reach the saturation magnetization according to the macroscopic measurements.

\subsection{Scattering experiment}
XRMS measurements were performed at T = 4.5~K on GdCo$_2$Zn$_{20}$ single crystal, mounted inside the closed-cycle displex cryostat in a six-circle diffractometer at the 6-ID-B beamline. 
The single crystal was oriented with the $[111]$ direction parallel to the vertical diffraction plane. 
Several magnetic superlattice reflections of the type $(\frac{L}{2}, \frac{L}{2}, \frac{L}{2})$ with $L = (2n+1)$ were measured and their integrated intensities were compared to the simulated intensities to determine the magnetic structure below $T_N$. 
To enhance the magnetic Bragg peak intensities, the energy of the incident beam was tuned near the Gd $L_2$ or $L_3$ absorption edges. 
In addition, in order to investigate any magnetic contribution from Co and Zn ions, the energy of the incident beam was also tuned to the Co (7709~eV) and Zn (9659~eV) $K$-edges and a search for superlattice reflections was performed. 
The charge and magnetic contributions present in the scattered beam were separated by a pyrolytic graphite [C$(006)$] analyzer crystal installed on the 2$\theta$ arm of the diffractometer. 
Since the incident beam presents the polarization perpendicular to diffraction plane ($\sigma$ polarization), by rotation of the analyzer crystal around the scattered beam wave vector $\vec{k'}$ we were able to select the two polarization channels ($\sigma-\sigma'$ and $\sigma-\pi'$) in this experimental geometry.\cite{hill1996}


\section{EXPERIMENTAL RESULTS}

The experimental results are organized into two sections: The first part is dedicated to the absorption measurements on powdered GdFe$_2$Zn$_{20}$ in its FM state. 
The subsequent section shows the results obtained by the XRMS technique on the GdCo$_2$Zn$_{20}$ single crystal in its AFM state. 

\subsection{Absorption measurements: GdFe$_2$Zn$_{20}$}

\begin{figure}[!b]
\centering
\includegraphics[trim=0.0cm 0.00cm 0.0cm 0.0cm, clip=true, totalheight=0.3 \textheight, angle=0]{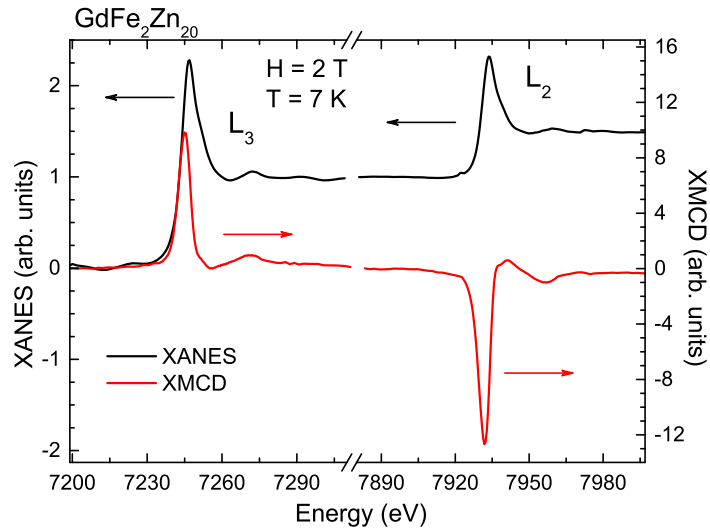}
\caption{(Color online) XANES and XMCD spectra obtained at the Gd $L_{2,3}$ edges performed at $T = 7$~K and $H$ = 2~T.}
\label{fig:fig1}
\end{figure}

XANES and XMCD measurements performed at the Gd $L_{2,3}$ edges in GdFe$_2$Zn$_{20}$ are shown in Fig.~\ref{fig:fig1}. 
Dipolar selection rules make the dichroic signal at the $L_{2,3}$ absorption edges sensitive to the spin polarization of the intermediate 5$d$ level.
The Gd XANES reported in Fig.~\ref{fig:fig1} are normalized to one at the $L_3$ and half at the $L_2$ edge to reflect the 2:1 ratio of the initial state at these edges (2$p_{3/2}$ and 2$p_{1/2}$, respectively). 
Figure~\ref{fig:fig1} also shows the XMCD spectra at the Gd $L_{2,3}$ edges in which each spectrum is normalized to the corresponding edge jump of the absorption spectrum. 
The XMCD signal obtained across the two edges show different intensities with a strong dichroic magnetic signal around 12.5~$\%$ at the $L_2$ and 9.7~$\%$ at the $L_3$ absorption edges, which is consistent with Gd-based compounds.\cite{Galera_JPCM_2008} 
The size and the shape of the magnetic contribution obtained by fitting the XMCD signals with Lorentzian function can describe additional properties of this system. 
The widths of the dipolar contributions (E1) contributions observed at the $L_2$ and $L_3$ absorption edges are 4.3(2)~eV and 4.5(2)~eV, respectively, which reflects a short 2$p$ core hole lifetime. 
Using the integrated intensities, the $L_3/L_2$ ratio (or Branching Ratio value - BR)\cite{Thole_PRB_1988, Laan_PRB_1990, Laan_PRB_1991, Laan_PS_1990, Veenendaal_PRL_1997} obtained experimentally is $-0.77(4)$.

\begin{figure}[!t]
\centering
\includegraphics[trim=0.0cm 0.00cm 0.0cm 0.0cm, clip=true, totalheight=0.3 \textheight, angle=0]{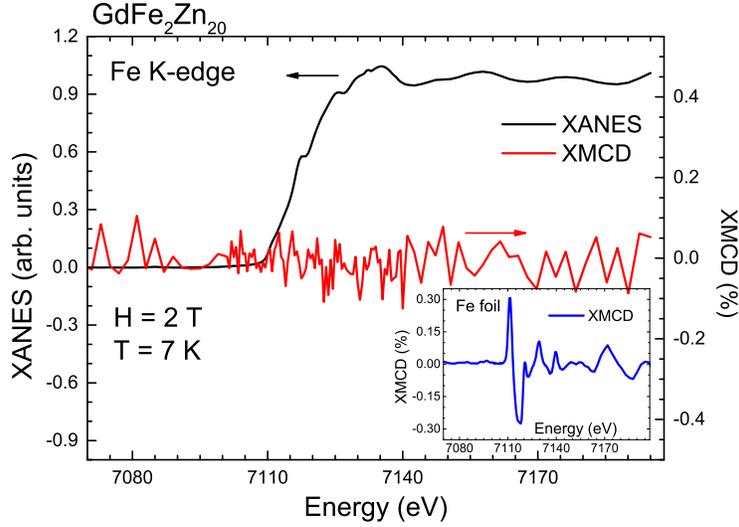}
\caption{(Color online) X-ray absorption measurements at 7~K and in 2~T at the Fe $K$-absorption edge for the GdFe$_2$Zn$_{20}$. The inset shows the XMCD data for the Fe-foil sample.}
\label{fig:fig1_n}
\end{figure}

\begin{figure}[!b]
\centering
\includegraphics[trim=0.0cm 0.00cm 0.0cm 0.0cm, clip=true, totalheight=0.3 \textheight, angle=0]{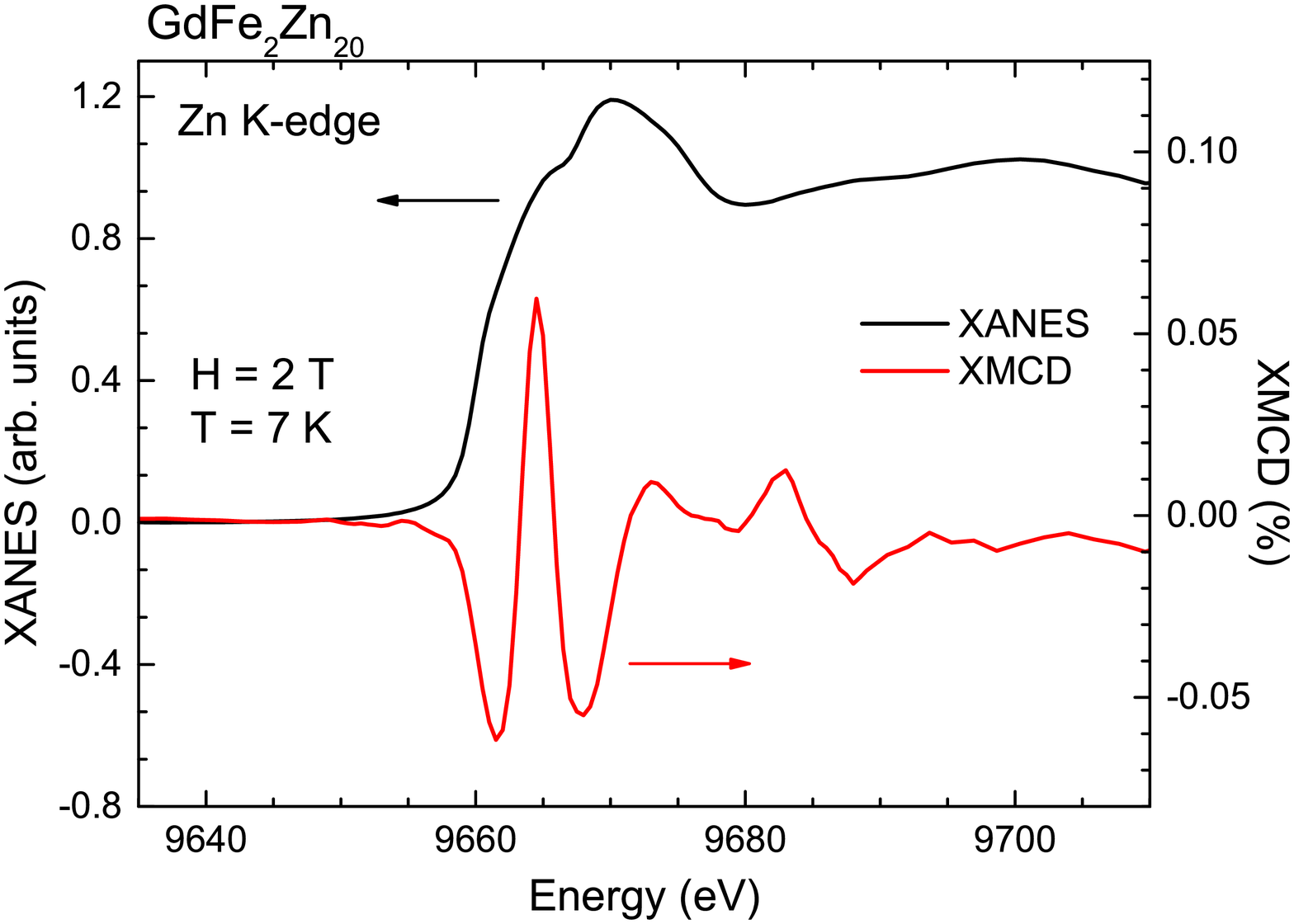}
\caption{(Color online) X-ray absorption measurements at 7~K and in 2~T for Zn $K$-absorption edge.}
\label{fig:fig2}
\end{figure}


Absorption measurements at the Fe and Zn $K$ edges were also carried out on powdered samples. 
The absorption measurement near the $K$ edge, in which the dipolar transition is probed (1$s \rightarrow$ 4$p$), is crucial towards understanding the magnetic and electronic properties due to the delocalized character of the $p$ states.\cite{Antonov2007}. 
Since the probed $p$ states are very delocalized, a strong influence of the surrounding matrix can be expected due to the hybridization between the rare earth and the transition metal ions. 
As shown in Figure~\ref{fig:fig1_n}, the measurements performed near the Fe $K$ edge do not reveal any magnetic contribution from the Fe ions higher than the background level ($\sim$ 0.07 $\%$). 
The inset in Fig.~\ref{fig:fig1_n} exhibits the XMCD measurements obtained for a 5 $\mu$m Fe-foil in the same experimental conditions: a clear dichroic signal can be observed near the edge.

However, the spectroscopy measurement at the Zn $K$ edge manifests an interesting behavior. 
Figure~\ref{fig:fig2} shows the absorption and dichroism results at the Zn $K$ edge in which an induced magnetic signal around 0.06~$\%$ is detected.  
This magnetic signal is due to a hybridization with the rare earth 5$d$ orbitals. 
The XMCD spectrum exhibits the main positive feature located at 9665~eV with a width around 2.5~eV, surrounded by two negative peaks 6.5~eV away. 
The broad feature localized around 9680~eV ($\sim$ 20~eV above the edge) is likely due to magnetic EXAFS.\cite{Rueff_PRB_1998}  

\begin{figure}[!tb]
\centering
\includegraphics[trim=0.0cm 0.00cm 0.0cm 0.0cm, clip=true, totalheight=0.45 \textheight, angle=0]{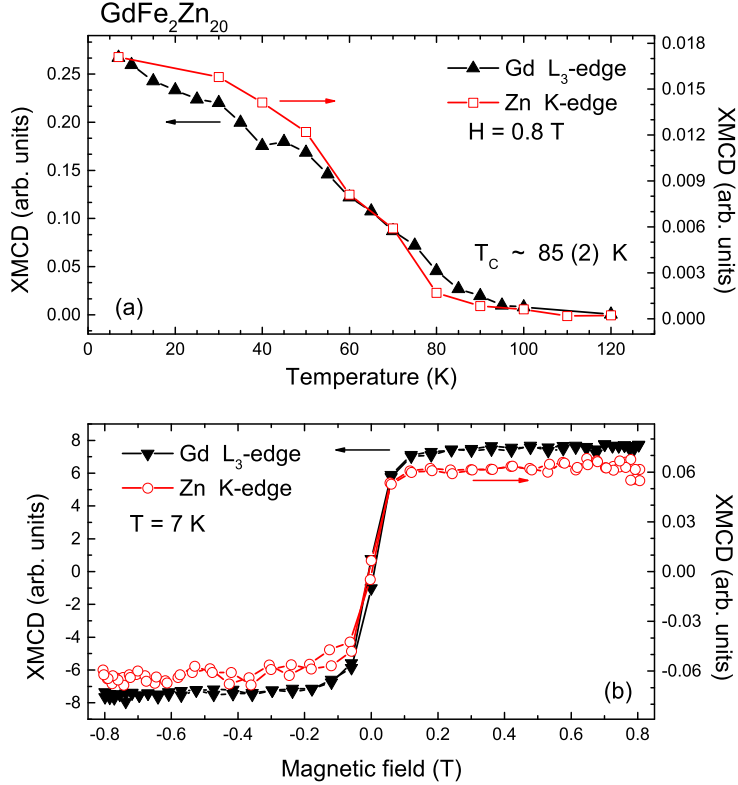}
\caption{(Color online) (a) Temperature dependence and (b) XMCD hysteresis loops obtained at the Gd $L_3$ and Zn $K$ - absorption edges. }
\label{fig:fig2tdep}
\end{figure}

A clear evidence of the Zn 4$p$ states polarization due to the Gd ions can be found in the temperature and field dependence reported in panels \ref{fig:fig2tdep}(a) and \ref{fig:fig2tdep}(b), respectively. 
The magnetic intensities for both Gd and Zn ions follow the same temperature evolution and disappear around the critical temperature (T$_C \sim$ 85~K).
In addition, the two hysteresis loops obtained at the maximum XMCD intensity shows clearly the Zn magnetism dependence in relation to the Gd ions and therefore it suggests a spin polarization of the Zn 4$p$ bands by the Gd sub-lattice.

\subsection{Magnetic scattering measurements: GdCo$_2$Zn$_{20}$}

\begin{figure}[!b]
\centering
\includegraphics[trim=0.0cm 0.0cm 0.0cm 0.0cm, clip=true, totalheight=0.3 \textheight, angle=0]{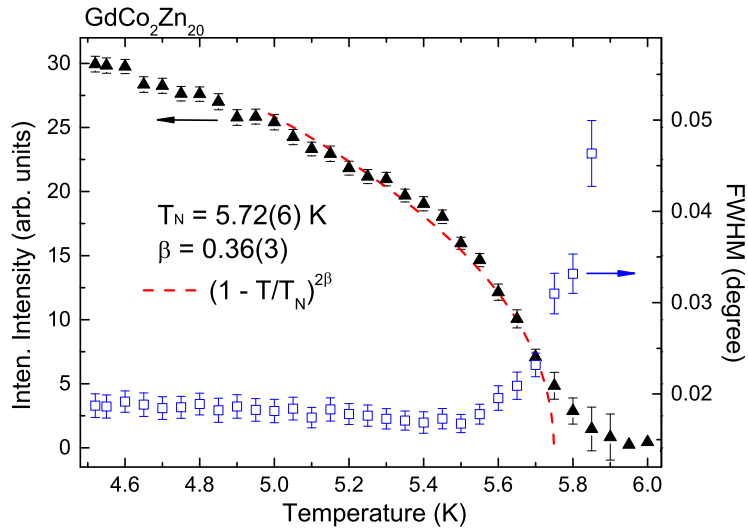}
\caption{(Color online) Temperature dependence of the integrated intensity measured with a rocking scan through the superlattice peak $(\frac{7}{2},\frac{7}{2},\frac{3}{2})$. The inset shows $T_N$ and the critical exponent $\beta$ obtained from a critical power-law fitting $(1-T/T_N)^{2\beta}$ of the intensities near $T_N$ (dashed red curve).}
\label{fig:fig3}
\end{figure}

Figure~\ref{fig:fig3} shows the evolution of the integrated intensity for the magnetic Bragg reflection $(\frac{7}{2},\frac{7}{2},\frac{3}{2})$ as a function of temperature for the GdCo$_2$Zn$_{20}$ compound fitted by a Lorentzian-Squared function. 
The magnetic peak intensity decreases smoothly to zero as the temperature approaches $T_N$, indicating a standard second order phase transition from an AFM to a paramagnetic state. 
A dashed red line in Fig.~\ref{fig:fig3} displays a fitting using a critical power-law expression, $(1-T/T_N)^{2\beta}$, above 5.0 K. 
The fitting around the N\'eel temperature yields a $T_N$ = 5.72(6) K and a critical exponent $\beta$ = 0.36(3). 
The value of $T_N$ is in good agreement with bulk magnetic susceptibility measurements and previous works.\cite{Jia_PRB_2008,  Jia_PRB_2009} 
The critical exponent $\beta$ close to 0.367 suggests a three-dimensional (3D) Heisenberg magnetic model.~\cite{Collins_Book_1989, Blundell_book_magnetism} 
In blue symbol (Fig.~\ref{fig:fig3}) is also reported the full width at half maximum (FWHM) of the magnetic superlattice peak $(\frac{7}{2},\frac{7}{2},\frac{3}{2})$ as a function of temperature. 
This figure clearly shows a peak broadening and a decrease in intensity near the phase transition temperature characteristic of a loss of long-range order. 

\begin{figure}[!t]
\centering
\includegraphics[trim=0.25cm 0.0cm 0.0cm 0.0cm, clip=true, totalheight=0.3 \textheight, angle=0]{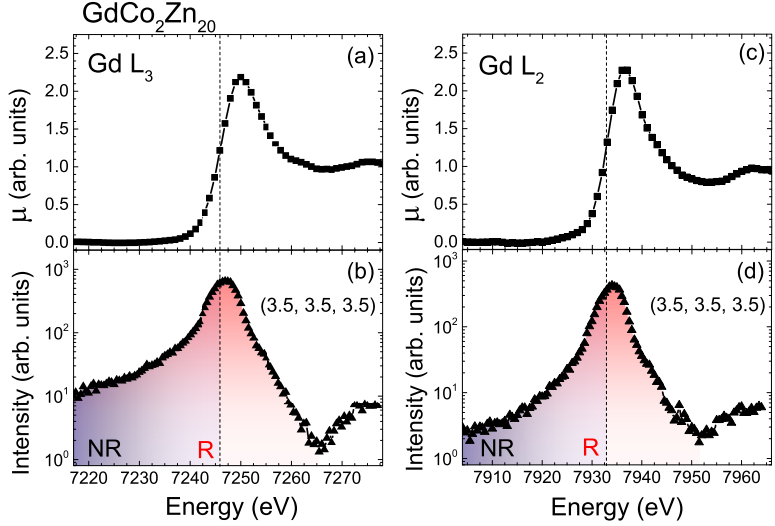}
\caption{(Color online) Energy dependences obtained across the Gd-$L_3$ and $L_2$ edges at 4.5~K. The upper pictures [(a) and (c)] display the absorption coefficients ($\mu$) obtained from fluorescence emission. The lower pictures [(b) and (d)] exhibit the energy dependences along the superlattice position $(\frac{7}{2}, \frac{7}{2},\frac{7}{2})$ in the $\sigma-\pi'$ polarization channel. The vertical dashed lines correspond to the position of the absorption edge inflection. NR and R stand for non-resonant and resonant sections in the (b) and (c) panels, respectively.}
\label{fig:fig4}
\end{figure}

Energy dependences across the Gd-$L_2$ and $L_3$ edges performed at 4.5~K are displayed in Fig.~\ref{fig:fig4}. 
The top panels [Fig.~\ref{fig:fig4}(a) and \ref{fig:fig4}(c)] show the normalized absorption coefficients ($\mu$) obtained from the fluorescence yield, while the bottom panels [Fig.~\ref{fig:fig4}(b) and \ref{fig:fig4}(d)] exhibit the energy profile obtained at the magnetic superlattice position $(\frac{7}{2},\frac{7}{2},\frac{7}{2})$ with the analyzer crystal set to the $\sigma-\pi'$ polarization channel. 
A resonant enhancement of over two orders of magnitude at both absorption edges can be seen. 
In addition, the maximum intensities are observed about 2-3 eV above the absorption edge (defined by the vertical dashed lines), which is a characteristic signature of a dipole electronic transition. 
The same energy dependence was performed in the $\sigma-\sigma'$ polarization channel, and no significant contribution was observed. 
The strong resonant enhancement in the spectra [Fig.~\ref{fig:fig4}(b) and \ref{fig:fig4}(d)] indicates a significant overlap between the initial 2$p$ and 5$d$ states, and a strong exchange interaction between the 4$f$-5$d$ orbitals.  
This magnetic polarization of the 5$d$ bands via 4$f$ states helps shed light on the magnetic structures of these rare earth based materials using the $L$ absorption edge measurements, i.e., 2$p \rightarrow$ 5$d$ transitions. 
Moreover, the asymmetric peak shape expressed as a long tail below the absorption edges arises from the interference between the resonant and non-resonant magnetic scattering contributions.\cite{Gibbs_1991, Kim_PRB_2005, Kim_EJP_2012}

\begin{figure}[!t]
\centering
\includegraphics[trim=0.1cm 0.0cm 0.0cm 0.2cm, clip=true, totalheight=0.4 \textheight, angle=0]{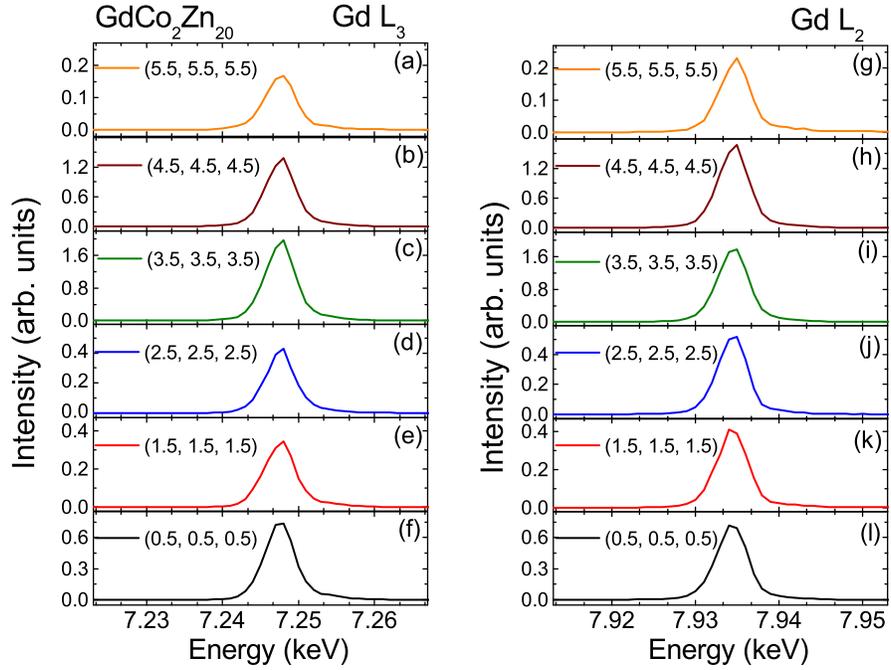}
\caption{(Color online) Energy dependence of the X-ray resonant magnetic scattering intensities for magnetic Bragg peaks of type $(\frac{L}{2},\frac{L}{2},\frac{L}{2})$ with $L = (2n+1)$ after absorption correction. (a)-(f): at the Gd $L_3$-absorption edge; (g)-(l): at the Gd $L_2$ edge. Both measurements were done in the $\sigma-\pi'$ polarization channel. }
\label{fig:fig5}
\end{figure}

The normalized energy line shape dependence after absorption correction for selected magnetic Bragg peaks $(\frac{L}{2},\frac{L}{2},\frac{L}{2})$ with $L = (2n+1)$ performed at Gd-$L_2$ and $L_3$ edges are displayed in Fig.~\ref{fig:fig5}. 
The magnetic reflections show a narrow resonant line shape at the $L_2$ ($w \le $ 4.9 eV) and $L_3$ ($w \le $ 5.9 eV) edges. 
The smaller energy broadening for the $L_2$ edge is associated with a relatively short lifetime compared with the $L_3$ edge.  
The magnetic reflections were only observed in the $\sigma-\pi'$ polarization channel; in the $\sigma-\sigma'$ channel, no intensity was detected. 
The integrated intensities of the magnetic reflections were used to extract the $L_3/L_2$ ratio and to determine the direction of magnetic moment. 
The BR values show an unusual behavior in which the values varies between $0.9$ - $1.2$ for different magnetic reflections. 
For XRMS, the branching ratio is expected to be equal to 1. 
The exact value for the BR ratio is difficult to obtain due to the various corrections that must be applied to the experimental data, such as self-absorption and angular corrections, and therefore, we cannot affirm that $\langle L^{5d}_z\rangle \not= 0$ for this AFM compound.

\subsection{Magnetic structure of GdCo$_2$Zn$_{20}$}

The magnetic structure of the Gd spins are determined comparing the experimental integrated intensities in Fig.~\ref{fig:fig5} with simulated data from selected magnetic reflections. 
The program SARA{\it h}~\cite{wills2000new} was run to determine the possible magnetic arrangements that the Gd ion can adopt inside the unit cell, i.e., the magnetic representation ($\Gamma_{Mag} $). 
In addition, we assumed that only the Gd ions carry magnetic moments in this compound. 
For this material, whose magnetic propagation vector is $(\frac{1}{2},\frac{1}{2},\frac{1}{2})$, the space group is $F d \bar{3} m$, and the Gd ions occupy the 8$a$ crystallographic site, the magnetic representation (MR) can be decomposed into four non-zero irreducible representations (IRs): two one-dimensional (1D - $\Gamma_{2,3}^{1}$) and two two-dimensional (2D - $\Gamma_{5,6}^{2}$) IRs. 
The four possible magnetic representations for the GdCo$_2$Zn$_{20}$ compound are summarized in~\autoref{tab:gamma_ir_bv}, where the labeling of the propagation vector and the magnetic representation follows the Kovalev notation.\cite{Kovalev_1993}

\begin{table}[!t]
\caption{Basis vectors (BV) for the space group $F d \bar{3} m$:2 with ${\bf k}_{9}=(\frac{1}{2},\frac{1}{2},\frac{1}{2})$. The decomposition of the magnetic representation (MR) for the Gd site can be written like $\Gamma_{Mag} =   \Gamma_{2}^{1} + \Gamma_{3}^{1} + \Gamma_{5}^{2} + \Gamma_{6}^{2}$. The two rare earth atoms positions of the nonprimitive basis are defined according to 1: $(0.125, 0.125, 0.125)$ and 2: $(0.875, 0.875, 0.875)$.}
\begin{tabular}{ccccccccccccccccccc}
\hline
\hline \noalign{\smallskip}
  IR  &  BV  &  Atom && \multicolumn{13}{c}{BV components}\\
 \cline{5-17} 
      &      &            &&&$m_{\|a}$ && $m_{\|b}$ && $m_{\|c}$ &&&$im_{\|a}$ && $im_{\|b}$ && $im_{\|c}$ \\
\hline \noalign{\smallskip}
$\Gamma_{2}$     & $\psi_{1}$ &      1 &&&                            1 &&                         1 &&      1 &&&      0 &&      0 &&      0  \\
                         &                &      2 &&&                            1 &&                          1 &&      1 &&&      0 &&      0 &&      0  \\
$\Gamma_{3}$    & $\psi_{2}$ &       1 &&&                            1 &&                         1 &&      1 &&&      0 &&      0 &&      0  \\
                         &                &      2 &&&                            -1 &&                        -1 &&     -1 &&&      0 &&      0 &&      0  \\
$\Gamma_{5}$    & $\psi_{3}$ &      1 &&&          $\nicefrac{1}{2}$ &&   -$\nicefrac{1}{2}$ &&        -1 &&&      0 &&      0 &&      0  \\
                        &                &       2 &&&         -$\nicefrac{1}{2}$ &&    $\nicefrac{1}{2}$ &&          1 & &&      0 &&      0 &&      0  \\
                        & $\psi_{4}$ &       1 &&&  0.866 &&  -0.866 &&     0 &&&      0 &&      0 &&      0  \\
                        &               &        2 &&&  -0.866&&   0.866 &&      0 & &&      0 &&      0 &&      0  \\
$\Gamma_{6}$ & $\psi_{5}$ &        1 &&&            $\nicefrac{1}{2}$ &&      $\nicefrac{1}{2}$ &&        -1 &&&      0 &&      0 &&      0  \\
                       &                &        2 &&&           $\nicefrac{1}{2}$ &&       $\nicefrac{1}{2}$ &&         -1 &&&      0 &&      0 &&      0  \\
                       & $\psi_{6}$ &        1 &&&  -0.866 &&   0.866 &&      0&&&      0 &&      0 &&      0  \\
                       &                &        2 &&&  -0.866 &&    0.866 &&      0&&&      0 &&      0 &&      0  \\
\hline
\hline \noalign{\smallskip}
\label{tab:gamma_ir_bv}
\end{tabular}
\end{table}

To determine the magnetic structure, the intensities were calculated assuming only dipole transition (E1) and hence, the X-ray magnetic scattering cross section model can be written as:\cite{hannon1988, hill1996, adriano2007magnetic, detlefs1997determination, nandi2011strong, mardegan2014}

\begin{eqnarray} \label{eq:eq2}
I_{RES} \propto  A * \left| \sum_{n} f^{E1}_{n}e^{i\vec{Q} \cdot \vec{R}_n}\right|^{2},
\end{eqnarray}
where
\begin{eqnarray} \label{eq:eq3}
 A = \frac{\sin(\theta+\alpha)\sin(\theta-\alpha)}{\sin \theta \cos \alpha \sin(2\theta)},
\end{eqnarray}
and
\begin{eqnarray} \label{eq:eq4}
\begin{split} 
 f^{E1}_{n} & =  [ (\hat{\varepsilon}'\cdot \hat{\varepsilon})F^{(0)} - i(\hat{\varepsilon}'\times \hat{\varepsilon}) \cdot \hat{z}_n F^{(1)} \\ 
& + (\hat{\varepsilon}'\cdot \hat{z}_n )(\hat{\varepsilon} \cdot \hat{z}_n)F^{(2)}  ] .
\end{split}
\end{eqnarray}

\noindent The term $A$ [Eq.~(\ref{eq:eq3})] contains the absorption correction and the Lorentz factor. 
$\alpha$ is the angle between the wave-vector transfer $\vec{Q} (=\vec{k'}-\vec{k})$ and the $[111]$ crystal direction, and $\theta$ is half of the 2$\theta$ scattering angle. 
The Eq.~(\ref{eq:eq4}) shows the resonant term. 
It carries information about the $\vec{k}$ ($\hat{\varepsilon}$) and $\vec{k'}$ ($\hat{\varepsilon}'$), i.e., the incident and scattered wave (polarization) vectors, respectively, and the magnetic moment direction at \textit{n}th site ($\hat{z}_{n}$). 
The terms $F^{(0,1,2)}$ are related to the dipole matrix transition and by atomic properties.\cite{hannon1988}
The exponential function in Eq.~\ref{eq:eq2} is a function of the wave-vector transfer $\vec{Q}$, and the position $\vec{R}_{n}$ of the \textit{n}th Gd ion inside the unit cell. 
For the XRMS technique probing AFM materials, we assumed that the magnetic intensity at the first harmonic satellites are due to only dipole contribution comes from the linear term on magnetic moment direction [$-i(\hat{\varepsilon}'\times \hat{\varepsilon}) \cdot \hat{z}_n$] displayed in Eq.~(\ref{eq:eq4}).

\begin{figure}[!t]
\centering
\includegraphics[trim=0.0cm 0.0cm 0.0cm 0.0cm, clip=true, totalheight=0.35 \textheight, angle=0]{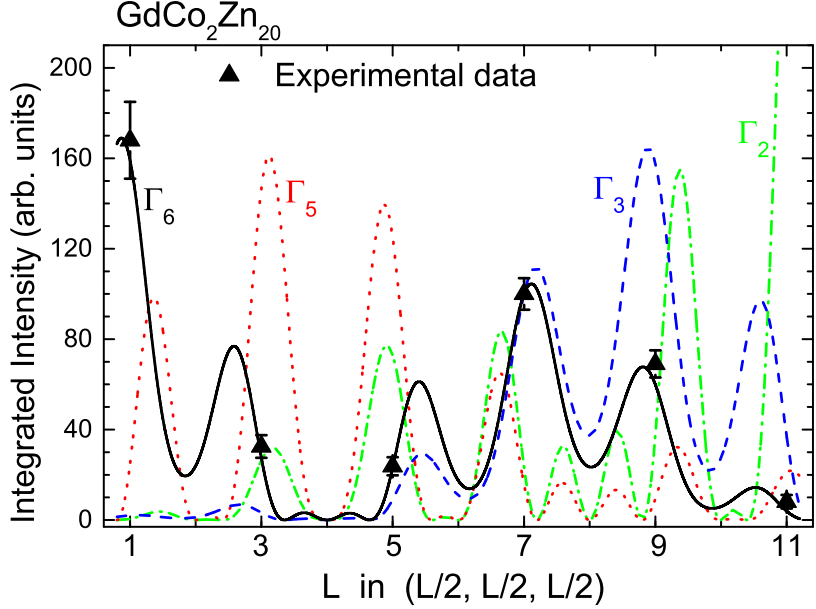}
\caption{(Color online) Simulated and experimental integrated intensities for magnetic reflections $(\frac{L}{2},\frac{L}{2},\frac{L}{2})$ with $L = (2n+1)$ performed at the Gd-$L_2$ absorption edge for GdCo$_2$Zn$_{20}$. The simulated intensities were obtained through Eq.~\ref{eq:eq2} supposing the four magnetic representations $\Gamma_{2}$, $\Gamma_{3}$, $\Gamma_{5}$ and $\Gamma_{6}$  summarized in~\autoref{tab:gamma_ir_bv}. The error bars for the experimental data were extracted from the fitting function.}
\label{fig:fig6}
\end{figure}

The simulated intensities obtained using Eq.~(\ref{eq:eq2})-(\ref{eq:eq4}) and the experimental intensities obtained at the Gd $L_2$ edge [Fig.~\ref{fig:fig4}(g-l)] are displayed in Fig.~\ref{fig:fig6}. 
Comparing the intensities, the best agreement between experimental (symbol, $\blacktriangle$) and simulated data (curves) is obtained for the magnetic representation $\Gamma_{6}$.

To identify different magnetic propagation vectors, such as $(0, \frac{1}{2},\frac{1}{2})$ and $(0, 0, \frac{1}{2})$, a systematic search for commensurate and incommensurate magnetic reflections in the reciprocal space was performed below $T_N$. 
However, only magnetic Bragg reflections of the type $(\frac{1}{2}, \frac{1}{2},\frac{1}{2})$ were observed. 
In addition, to probe a possible presence of AFM magnetic moments in the Co and Zn ions, the beam energy was tuned to the Co and Zn $K$ edges. 
Search for magnetic superlattice reflections were performed below $T_N$, and no measurable magnetic reflection at 4.5 K could be observed.

\section{DISCUSSION}

The nature of the electronic and magnetic properties of the intermetallic $RT_2$Zn$_{20}$ systems depends strongly on the interactions between the rare earth and the transition metals ions. 
Since the Gd 5$d$ states participate on the conduction bands, the resonant absorption and diffraction measurements at the Gd $L_{2,3}$ edges provide valuable information. 
As reported in susceptibility measurements and band structure calculations\cite{Jia_Nature_2007,Jia_PRB_2008}, the replacement of the transition metal affects the electronic density of states at the Fermi level ($\rho_{E_F}$) and the conduction band without changing significantly the lattice parameters.\cite{lattice} 
Doping studies of Gd(Fe$_x$Co$_{1-x})_2$Zn$_{20}$ compounds shows a monotonic increase of the magnetic ordering temperature when $x$ increases to 1 which indicates that there is an increase in the coupling between the rare earth magnetic moments.\cite{Jia_Nature_2007}

Absorption measurements performed in powdered samples of FM GdFe$_2$Zn$_{20}$ compound below $T_C$ reveal interesting behaviors. 
As can be seen in Fig.~\ref{fig:fig1}-\ref{fig:fig2tdep}, only at the Gd $L$ and Zn $K$-edges a dichroic signal is observed above the background level whilst an unexpected lack of magnetic intensity is observed at the Fe $K$-edge. 
The intense magnetic signal at the rare earth $L$-edges occurs mainly due to the overlap between the Gd 2$p$ and 5$d$ states and a strong energy splitting of the 5$d$ sub-bands as a result of a 4$f$-5$d$ exchange interaction.\cite{Harmon1974spin,Harmon1974spin2, Koenig_JAP_1994, Book_MSR_2006}
In addition, the splitting of the $d$ state into 5$d$ spin-up and spin-down wave function has considerable influence in the magnetism observed at the Zn $K$-edge. 
The Zn 3$d$ orbitals are completely filled (3$d^{10}$) and henceforth, a magnetic moment due to an overlap between the 3$d$ and 4$sp$ orbitals in the Zn ion is not expected to occur due to the filled 3$d$ orbitals being more contracted. 
Therefore, the magnetic signal observed in the 4$p$ states is due to hybridization with the extended Gd 5$d$ orbitals and not from the exchange interaction with the Zn 3$d^{10}$ orbitals.\cite{Rueff_PRB_1998, Galera_JPCM_2008} 
Following Hund's rule, the Gd ion has the 4$f$ state filled by seven spin-up electrons in which it pulls the 5$d$ sub band spin-up function towards the inner core due to a positive exchange interaction. 
The short distances between the first Gd-Zn ions ($\sim$ 3 \AA{}) drives a small hybridization between the broad Zn 4$p$ and Gd 5$d$ states, inducing a small amount of magnetic moment in 4$p$ states.

Theoretical works have suggested that the orbital moment should be almost zero for the 5$d$ band ($\langle L^{5d}_z\rangle = 0 $), i.e., a quenching of the angular momentum, so the 5$d_{3/2}$ and 5$d_{5/2}$ sub-bands should display the same polarization and thereby the dichroism at the $L_3$ and $L_2$ edges should have equal magnetic intensity.\cite{Harmon1974spin,Veenendaal_PRL_1997, Imada_RMP_1998} 
For the GdCo$_2$Zn$_{20}$ compound, the BR ratio vary between 0.9 - 1.2, i.e., close to the theoretical value and therefore we cannot suggest any orbital moment for this compound. 
However, as showed in Fig.~\ref{fig:fig1} for the FM compound, the intensity recorder at the $L_2$ is higher than the $L_3$ edge, where we observe a $L_3/L_2$ ratio of approximately $-0.77(4)$.\cite{SO_minus}  
This slightly different value from the theoretical branching ratio expected for XMCD (BR $= -1$) suggests that the Gd ions may carry a small orbital moment at the 5$d$ orbitals in the FM compound.

M\"{o}ssbauer spectroscopy measurements\cite{Tamura2013study} and simulations\cite{Isikawa2013enhancement} for the DyFe$_2$Zn$_{20}$ compound reported that the Fe ions align AFM with the Dy magnetic moments in which the iron ions exhibit a very small magnetic moment $\sim$ $0.2~\mathrm{\mu_B}$/Fe. 
Neutron diffraction measurements\cite{tian2010magnetic} performed in TbFe$_2$Zn$_{20}$ compound also reported that a small magnetic moment at the Fe ions ($< 1~\mathrm{\mu_B}$) would improve the refinement. 
In addition, recently M\"{o}ssbauer measurement\cite{Budko_2015} in GdFe$_2$Zn$_{20}$ also reported the presence of a small magnetic contribution at the Fe site. 
In order to verify the quality of our XMCD data at the Fe $K$ edge, dichroic measurements at low temperature and under magnetic field in a 5 $\mu$m Fe-foil was also carried out. 
A magnetic signal around $0.3~\%$ which, according to the literature\cite{Stahler_PRB_1993, Iota_APL_2007}, corresponds to a magnetic moment around $2.2~\mathrm{\mu_B}$/Fe was observed.
Assuming that Fe spins order ferromagnetically in GdFe$_2$Zn$_{20}$ with a magnetic signal below 0.07 $\%$ (noise level), it would result in a magnetic moment lower than $\mathrm{\mu_{Fe}} \sim 0.5~\mathrm{\mu_B}$, i.e., the same order of magnitude as reported by band structure calculation\cite{Jia_Nature_2007} and comparable with the DyFe$_2$Zn$_{20}$ compound. 
Therefore, if the Fe ions carry magnetic contribution in the GdFe$_2$Zn$_{20}$ materials, we can suggest that this magnetic moment must be lower than $ 0.5~\mathrm{\mu_B}$. 
One possible explanation for the difficulty in identifying the magnetic signal at the Fe $K$-edge (7112 eV) would be due to contamination coming from the Gd $L_3$ edge (7243 eV). 
Despite the fact that the two edges are apart by around 130 eV, the Gd $L_3$ pre-edge increases the background around the Fe $K$-edge which, as a consequence may hide the small magnetic signal.  
An AFM ordering state for the Fe ions cannot be ruled out, but is less likely to occur.

Field dependent magnetization measurements\cite{Jia_Nature_2007} show that the GdFe$_2$Zn$_{20}$ compound reaches a saturation magnetization of 6.5 $\mathrm{\mu_B}$/f.u. at low temperature. 
Assuming that the Gd ions contribute with 7.94 $\mathrm{\mu_B}$/Gd for the total magnetic moment (theoretical value), and that the transition metals are coupling antiferromagnetically with the rare earth elements, the transition metals are found to carry a magnetic moment around 1.44 $\mathrm{\mu_B}$/f.u. opposite to the Gd ions. 
Supposing that the Fe ions have a total magnetic moment $\mathrm{\mu_{Fe}} \sim 0.5~\mathrm{\mu_B}$, consequently the Zn ions present in this material would present an induced magnetic moment of approximately 0.05 $\mathrm{\mu_B}$/Zn. 
Therefore, it strongly suggests that the interaction between the rare earth ions affects the environment around the atoms and consequently spin polarizes the transition metal ion. 
The rough estimate of magnetic moment for the Zn ions based on spectroscopy and macroscopic measurements has to be further investigate by density functional theory calculation.

To provide further information about these systems, the magnetic properties of the AFM compound GdCo$_2$Zn$_{20}$ were also investigated, using the XRMS technique rather than neutron scattering, due to the high neutron absorption cross-section of Gd ions. 
The transition to the magnetically ordered phase driven by temperature is characterized by the appearance of superlattice magnetic reflections with a magnetic propagation vector $\tau = (\frac{1}{2}, \frac{1}{2},\frac{1}{2})$. 
This $\tau$ magnetic vector indicates that the magnetic unit cell is represented by a doubled chemical unit cell in all three crystallographic directions.
As displayed in Fig.~\ref{fig:fig5}, the Gd spins follow the magnetic representation $\Gamma_{6}$ which is different when compared with the isostructural TbCo$_2$Zn$_{20}$ compound.\cite{tian2010magnetic} 
Since there is a large separation ($\sim$ 6 \AA{}) between the rare earth ions in this system, the magnetic properties will be mainly mediated via the conduction electrons. 
The large distance between the first $R$-$R$ ions explains quite well a weakening of the $J_{\mathrm{RKKY}}$ exchange interaction and thus a very low representative magnetic transition temperature, i.e., approximately 5.7~K (GdCo$_2$Zn$_{20}$) and 2.5~K (TbCo$_2$Zn$_{20}$). 
As a consequence of a poor $J_{\mathrm{RKKY}}$ coupling, the surrounding matrix or the Zn cages around the $R$ ions are weakly affected and hence it is not possible to induce magnetic moment in the transition metal ions. 
The absence of magnetism in the transition metal ions are also observed in the TbCo$_2$Zn$_{20}$ compound, in which the best refinement of the magnetic structure was obtained with null magnetic moment in the Co and Zn ions. 
Although GdCo$_2$Zn$_{20}$ and TbCo$_2$Zn$_{20}$ compounds display the same magnetic propagation vector $(\frac{1}{2},\frac{1}{2},\frac{1}{2})$, their magnetic structures are different, which is mainly related to the competition between the RKKY and CEF interactions. 
For the Tb-based materials the crystalline electric field splitting can affect the $J_{\mathrm{RKKY}}$ constant and therefore influence the magnetic coupling and the temperature transition.  
It is well known that an energy level CEF splitting induces magnetic anisotropies in the ground state and it may influence the total angular momentum. 
Several Tb-based intermetallics have exhibited distinct magnetic properties in relation to the Gd counterparts as shown in the layered family $R_nT_mM_{3n+2m}$ ($R$ = rare earth; $T$ and $M$ = transition metal; $n$ = 1,2 and $m$ = 0,1). \cite{Thompson2012,Thompson2001,Coleman2008} 
Although GdCo$_2$Zn$_{20}$ and TbCo$_2$Zn$_{20}$ compounds display the same magnetic propagation vector $(\frac{1}{2},\frac{1}{2},\frac{1}{2})$, their magnetic structures are different, which is mainly related to the competition between the RKKY and CEF interactions. 
The modification in the magnetic structure for different rare earth ions but with the same propagation vector has already been extensively investigated and it is mainly related to the RKKY and CEF interactions. 
Jia \textit{et al}.[\onlinecite{Jia_PRB_2009}] evaluated the CEF parameters from the thermodynamic measurements for the entirely $R$Co$_2$Zn$_{20}$ series ($R$ = Tb-Tm) and they observe small energy scales (small $W$ parameter) and a large $B_6^0$ CEF parameter for the complete series. 
This finding suggests a small energy level splitting and a strong influence of the Zn cage on the rare earth ions, i.e., guest-framework interaction. 
Therefore, we can suggest that the rare earth ions located in this large polarized environment are strongly affected by the Zn cages, which has direct influence in the electronic and magnetic properties. 
This can be seen in the different magnetic structures for the AFM compounds and the spin polarization of the Zn ions only for the GdFe$_2$Zn$_{20}$ compound. 
Hence, the CEF effect has an important role in this class of compound.  
A detailed investigation for different rare earth elements would allow a better understanding of the $R$Co$_2$Zn$_{20}$ family. 
Nevertheless macroscopic measurements up to 1.8 K only report magnetic ordering for the compounds with $R$ = Gd and Tb.


\section{SUMMARY}

We have investigated the intermetallic Gd$T_2$Zn$_{20}$ system with $T =$ Co and Fe at low temperature using the XRMS and XANES/XMCD techniques, respectively. 
The XRMS measurements performed in GdCo$_2$Zn$_{20}$ compound reveal a commensurate antiferromagnetic ordering with a magnetic propagation vector $(\frac{1}{2},\frac{1}{2},\frac{1}{2})$ in which only the Gd ions carry magnetic moments. 
Selected magnetic reflections were measured in the polarization channel $\sigma-\pi'$ and we identified that the Gd spins follow the magnetic representation $\Gamma_6$, which is different from the isostructural compound TbCo$_2$Zn$_{20}$, mainly due to the CEF effects in the latter. 
The evolution of magnetic signal showed a magnetic phase transition below T$_N$ = 5.72(6)~K with a critical exponent $\beta$ = 0.36(3), suggesting a three-dimensional (3D) Heisenberg magnetic model. 
The XANES and XMCD measurements performed at the Gd $L_{2,3}$ edges in GdFe$_2$Zn$_{20}$ compound reveal a strong magnetic signal ($\sim$ 12.5~$\%$ - $L_2$ and 9.7~$\%$ - $L_3$) indicating a splitting of the 5$d$ orbitals and a strong Gd-Gd exchange interaction as well as a non zero orbital moment. 
In addition, we observed a presence of a small magnetic dichroic signal at the Zn $K$ edge due to the spin polarization of the Gd 5$d$ orbitals. 
This indicates a large RKKY exchange interaction between the Gd-Gd ions which polarizes the surrounding matrix.

\section*{acknowledgments}
This work was supported by FAPESP (SP-Brazil) under Contracts No. 2009/10264-0, 2011/19924-2, 2011/24166-0, 2012/10675-2 and 2012/17562-9. 
Parts of this research were carried out at the light source PETRA III at DESY, a member of the Helmholtz Association (HGF). 
Work at Argonne is supported by the U.S. Department of Energy, Office of Science, Office of Basic Energy Sciences, under Contract No. DE-AC-02-06CH11357. 
We would like to thank R. D. Reis and C. S.B. Dias for their assistance in macroscopic measurements and J. W. Kim for his assistance at beamline 6-ID-B. 
We are indebted to D. Reuther, R. Kirchhof  and H.-C. Wille for their assistance with the electromagnet at P09/DESY. 


\begin{thebibliography}{10}
\bibitem{lattice} The GdFe$_2$Zn$_{20}$ compound has a cubic lattice parameter $a$ around 0.07 \AA{} larger than GdCo$_2$Zn$_{20}$. Hence, a modification in the unit cell volume $\sim3\%$ by replacing the ion.
\end{thebibliography}




\end{document}